\documentstyle[tighten,preprint,eqsecnum,aps,prd]{revtex}
\begin{document}
\draft
\baselineskip=16pt   
\hyphenation{mani-fold mani-folds geo-me-try mi-ni-su-per-spa-ce
pro-ba-bi-lis-tic ge-ne-ral par-ti-cu-lar
}

\title{ Quantum Cosmology in  Bergmann-Wagoner Scalar-tensor 
        Gravitational Theory }

\author{Luis O. Pimentel\footnote{ E-mail: lopr@xanum.uam.mx}}

\address{ Departamento de F\'{\i}sica, Universidad Aut\'onoma Metropolitana, Iztapalapa
\\
P.O. Box 55-534,
CP 09340, M\'exico D. F., MEXICO }

\author{C\'esar Mora \footnote{E-mail: ceml@xanum.uam.mx}}

\address{ Departamento de F\'{\i}sica,
Universidad Aut\'onoma Metropolitana, Iztapalapa
\\
P.O. Box 55-534,
CP 09340, M\'exico D. F., MEXICO
\\
and
\\
Departamento de F\'{\i}sica,
UPIBI-Instituto Polit\'ecnico Nacional,\\
Av. Acueducto s/n Col. Barrio La Laguna Ticom\'an, \\
CP 07340 M\'exico DF, MEXICO.}

\date{\today}
\maketitle
\begin{abstract}
The Wheeler-DeWitt equation is solved for the Bergmann-Wagoner
scalar-tensor gravitational theory in the case of Friedmann-Robertson-
Walker cosmological model. We present solutions for several 
cosmological functions: $i) \,\lambda(\phi)=0, \,\,ii)\, \lambda(\phi)=
3\Lambda_0\phi$ \,and \, $iii) \,$ a more complex $\lambda(\phi)$, that depends on the choice of the coupling function, considering closed, flat and hyperbolic Friedmann universes ($k=1, 0, -1$). In the first two cases we show particular quantum wormhole solutions. Also, classical solutions are considered for some scalar-tensor theories, and we study the third quantization of some minisuperspace models.
\end{abstract}

\narrowtext


\section{Introduction}

In this work we will consider the quantum cosmology of the Bergmann-Wagoner (BW) theory of gravitation\cite{BerWag} in the homogeneous and isotropic case. This theory is the most general scalar-tensor theory of gravitation. The action for this theory contains two arbitrary functions of the scalar field, $\omega(\phi)$ and $\lambda(\phi)$. Every specific choice of these functions define a particular scalar-tensor theory. The function $\omega(\phi)$ is a coupling function. A dual role is played by the function $\lambda(\phi)$, it acts as a dynamical cosmological function that can give cosmological models with small cosmological constant at present times consistent with the observations but large at earlier times to produce inflation, and is also a potential function in the equation for the scalar field. Among the particular cases of BW theory we have: (1) Brans-Dicke theory ($\omega(\phi)= \omega_0=const.,\lambda(\phi)=0$ )\cite{Brans-Dicke}, (2) low energy limit of string theory ($\omega(\phi)=-1$), Barker's constant $G$ theory ($\omega(\phi)=(4-3\phi)/(2\phi -2),\lambda(\phi)=0$ )\cite{Barker}, Kaluza-Klein theory ($\omega = -(n-1)/n$, where the space-time dimension is $4+n$)\cite{Cho}. In the past several cosmological consequences of the classical BW theory have been studied: exact solutions, qualitative analysis of the equations of motion and the implementation of inflation ("no-hair" theorem).

 Very little attention has been paid to the quantum cosmology of this general theory,  however several particular case has been studied\cite{Gasperini}.  Therefore it seems of interest to consider the quantum cosmology obtained from the BW theory. In what follows we consider the quantum cosmology  of BW theory, that is, we keep as far as possible $\omega(\phi)$ and $\lambda(\phi)$ arbitrary in the homogeneous and isotropic line element, we obtain the Wheeler-DeWitt(WDW) equation\cite{DeWitt} for this model. When we try to solve exactly the WDW equation by separation of variables, it is found that we can do so for arbitrary $\omega(\phi)$, but only for particular forms of $\lambda(\phi)$. We solve the WDW equation for several potentials, finding general solutions to the separated equations and from them we can use superposition to obtain wavefunctions that satisfy particular boundary conditions. In particular we consider the wormhole wavefunctions that satisfy the HP regularity conditions\cite{Hawking1}: {\em{(i)}} the wavefunction is exponentially damped for large spatial geometry, and {\em{(ii)}} the wavefunction should be well behaved (regular) at the origin. This is because the wavefunction should represent Euclidean space for large radius and there should be no singularities when the spatial geometry degenerates. At the classical level, BW theory is known to have solutions representing wormholes\cite{Xiao}. Then, we explore the possibility to get wormholes in the quantum case. 

From the field equations one can also consider the classical problem; for those cases in which classical solutions were not known, they are given. Explicitely for Brans-Dicke theory, Barker's theory and three families of other theories that have been studied previously\cite{Barrow1,Barrow2}. In one particular case we obtained classical non-singular solutions. 

On the other hand, because the fact that the WDW equation is a second-order hyperbolic differential equation, there is no conserved positive-definite probability density as in the case of the Klein-Gordon equation. Then, there is a problem in the interpretation that $\mid \psi \mid^2$ is a probability, where $\psi$ is a solution to the WDW equation. An alternative approach is the third quantization procedure\cite{Hosoya}, in analogy with the second quantization of the Klein-Gordon equation. Thus, it can be construct a consistent probability measure in quantum gravity by promoting $\psi$ to a quantum field operator that acts on a Hilbert space of states. Then, third-quantized universe theory describes a system of many universes.
For one of the potentials and the closed universe we carry out the third quantization of the model and calculated the number of universes produced from nothing. 

Another interesting aspect on the sense of third quantization is the study of quantum effects in the universe evolution, for a specific  potential we calculated the uncertainty relations and found that the quantum fluctuations are large for small  values of the expansion factor of the universe and are bounded for large values of it.                                                                                                       

The organization of the paper is as follows: in the Sections II-IV we give a short review of the minisuperspace models that lead to the class of WDW equations considered here, and respective quantum solutions are given. In Sec. V we show a collection of classical solutions for some scalar-tensor theories. In Sec. VI the quantum wormhole configurations  are studied, Sec. VII is devoted to study the third quantization of particular minisuperspace model for $N=1/a\phi^{\frac{1}{2}}$. In Sec. VIII we calculate the fluctuations of a third quantized minisuperspace model for $N=1$. Finally, we summarize and conclude in Sec. IX. 


\section {The Bergmann-Wagoner minisuperspace model}

In order to construct the minisuperspace model, our starting point is 
the action for Bergmann-Wagoner scalar-tensor theory:
\begin{equation}
S=\frac{1}{l_p^2}\int_M\!\sqrt{-g}\left[\phi R^{(4)} -
\frac{\omega(\phi)}{\phi} g^{\mu\nu}\phi_{,\mu}\phi_{,\nu}+
2\,\phi\,\lambda(\phi)\right]\,d^4\!x 
+\frac{2}{l_p^2}\int_{\partial
M}\sqrt{h}\phi
h_{ij}K^{ij}\,d^3x, 
\label{action1}
\end{equation}
where $g=det(g_{\mu,\nu}), R^{(4)}$ is the scalar curvature  
of the Friedmann-Robertson-Walker(FRW) theory, $\phi(t)$ is the conventional real scalar gravitational field, $l_p$ is the Planck length and $\lambda(\phi)$ is the cosmological term. The second integral is a 
surface term involving the induced metric $h_{ij}$ and second 
fundamental form $K^{ij}$ on the boundary, needed to cancel the second 
derivatives in $R^{(4)}$ when the action is varied with the metric and 
scalar field, but not their normal derivatives, fixed on the boundary. 
We want to study an homogeneous and isotropic cosmological 
model, consequently we used the FRW metric line element in spherical polar coordinates $(t,r,\theta,\Phi)$, given by
\begin{equation}
ds^2=-N^2(t)dt^2 + a^2(t)\left[\frac{dr^2}{1-kr^2}+
r^2(d\theta^2+\sin^2{\theta}\,d\Phi^2)\right],
\label{metricFRWa}
\end{equation}
where $N$ is the lapse function, $a$ is the scale factor of 
the universe  and the Ricci scalar is
\begin{equation}
R^{(4)}=-\frac{6k}{a^2}-6\frac{\dot
a^2}{N^2a^2}- 6\frac{\ddot a}{N^2a} + 6\frac{\dot a\dot N}{N^3a},
\label{Ricci}
\end{equation}
substituting Eq. (\ref{Ricci}) into Eq. (\ref{action1}) and integrating with respect to space coordinates, after simplifications we have
\begin{equation}
S=\frac{1}{2}\int\!\left[-Nka\phi + \frac{a\phi}{N}\dot a^2 + 
\frac{a^2}{N}\dot a\dot \phi
-\frac{N\omega(\phi)}{6\phi}a^3\dot\phi^2 + 
\frac{N}{3}a^3\phi\lambda(\phi)
\right]dt,
\label{action2}
\end{equation}
where dot denotes the time derivative with respect to the time $t$, now we introduce the following new variables
\begin{equation}
x = a\phi^\frac{1}{2} , \quad  y = \int\left(\frac{2\omega(\phi)+3}
{12}\right)^\frac{1}{2}\frac{d\phi}{\phi} ,
\quad  \Lambda(y) =  \frac{\lambda(\phi)}{3\phi} , \quad
d\tau = \phi^\frac{1}{2}dt , 
\label{variables1}
\end{equation}
then the BW action simplifies to
\begin{equation}
S_{BW} = \frac{1}{2}\int\left[\frac{x}{N}x^{\prime 2} - 
\frac{x^3}{N}y^{\prime 2} - Nkx + Nx^3\Lambda(y) \right]d\tau,
\label{action3}
\end{equation}
prime denotes the time derivative with respect to the new time $\tau$. The lapse function $N$ is a Lagrangian multiplier whose variation leads to a constraint equation. The Hamiltonian can be constructed according to the standard canonical rule, and we expressed it as $H=N{\cal H}$. Since $N$ is a Lagrange multiplier, we have the constraint ${\cal H}\approx0$, and follow the Dirac quantization procedure; the wavefunction of the universe is annihilated by the Hamiltonian operator  ${\hat{\cal H}}\psi=0$, this equation is know as WDW equation. We can see that this wave equation is independent of the lapse, however in the following sections we study particular gauges $N=1$ and $N=1/x$, that results in separable WDW equations. 


\section{ Gauge N=1}

With the choice $N=1$, the action (\ref{action3}) becomes
\begin{eqnarray}
S_{BW} &=& \frac{1}{2}\int\left[xx^{\prime 2} - 
x^3y^{\prime 2} - kx + x^3\Lambda(y) \right]d\tau, \nonumber \\
&\equiv& \int L\,d\tau,
\label{action4}
\end{eqnarray}
hence the canonical conjugate momenta corresponding to $x$ and $y$ are given by
\begin{equation}
\pi_x = \frac{\partial L}{\partial x^{\prime}} = 
xx^{\prime}, 
    \quad 
\pi_y = \frac{\partial L}{\partial y^{\prime}} = 
-x^3y^{\prime}.
\label{momenta1}
\end{equation}
The Hamiltonian $H$ of the system is
\begin{equation}
{\cal H} = \frac{1}{2}\left[ x^{-1}\pi^2_x - x^{-3}\pi^2_y + kx - 
x^3\Lambda(y) \right],
\label{hamiltonian1}                       
\end{equation}
now, the canonical momenta in Eq. (\ref{hamiltonian1}) are converted into operators in the standard way, $\pi_x^2 \to -x^{-p}\frac{\partial}{\partial x}\left(x^p\frac{\partial}{\partial x}\right)$  and  $\pi_y^2 \to -\frac{\partial^2}{\partial y^2}$, the ambiguity of factor ordering is encoded in the $p$ parameter, then the WDW equation  is
\begin{equation}
\left[x^2\frac{\partial^2}{\partial x^2} + px\frac{\partial}{\partial x} - \frac{\partial^2}{\partial y^2} 
- kx^4 + x^6\Lambda (y)\right]\psi(x,y)=0.
\label{wdw1}
\end{equation}
It is easy to see that this wave equation can be solved by means of separation variables if $\Lambda(y)=const.$, thus in the next subsections we present three simple solvable cases for different cosmological terms.


\subsection{Case $\Lambda(y)=0$}

First, let us consider the WDW equation without a cosmological term, then  we have 
\begin{equation}
\left[x^2\frac{\partial^2}{\partial x^2} + px\frac{\partial}{\partial x} - \frac{\partial^2}{\partial y^2} -kx^4 \right]\psi(x,y)=0,
\label{wdw2}
\end{equation}
in order to solve this wave equation we make $\psi(x,y) = e^{isy}X(x)$, thus we obtain the following  solution given in terms of Bessel functions. In the next, we show the corresponding universe wavefunction for the three possible values of spatial curvature constant $k=-1, 0, 1$.
\begin{flushleft}
\noindent  {\em i) $k=1$} 
\end{flushleft}
In a closed FRW model we have that the universe wavefunction is
\begin{equation}
\psi_s(x,y) = x^{\frac{1-p}{2}}y^{\frac{1}{2}}
\left[C_1I_m\left(\frac{x^2}{2}\right) +
C_2K_m\left(\frac{x^2}{2}\right)\right]
\left[C_3e^{isy} + C_4e^{-isy}\right].
\label{sol1}
\end{equation}
\begin{flushleft}
\noindent  {\em ii) $k=-1$} 
\end{flushleft}
In the hyperbolic FRW universe, the solution is expressed in terms of modified first kind Bessel functions 
\begin{equation}
\psi_s(x,y) = x^{\frac{1-p}{2}}y^{\frac{1}{2}}
\left[C_1I_m\left(\frac{x^2}{2}\right) + 
C_2I_{-m}\left(\frac{x^2}{2}\right)\right]\left[C_3e^{is y} + C_4e^{-is y}\right],
\label{sol7}
\end{equation}
where $s$ is a separation constant, and the Bessel function order is given by $m = \frac{1}{2}\sqrt{\left(\frac{1-p}{2}\right)^2 - s^2}$. \begin{flushleft}
\noindent  {\em iii) $k=0$} 
\end{flushleft}
For a flat FRW universe the wavefunctions $\psi(x,y) = e^{isy}X(x)$ are
\begin{flushleft}
\noindent  {\em a) $ s=(1-p)/2$} 
\end{flushleft}
\begin{equation}
\psi_s(x,y)=x^{\frac{1-p}{2}}y^{\frac{1}{2}}
\left[C_1+C_2\ln{x}\right]\left[C_3e^{is y} + C_4e^{-is y}\right],
\label{sol5}
\end{equation}
\begin{flushleft}
\noindent  {\em b) $s\neq (1-p)/2$} 
\end{flushleft}
\begin{equation}
\psi_s(x,y)=x^{\frac{1-p\mp\sqrt{(1-p)^2-4s^2}}{2}} y^{\frac{1}{2}}
\left[C_1+C_2x^{ \pm \sqrt{(1-p)^2-4s^2}}\right]\left[C_3e^{is y} + 
C_4e^{-is y}\right].
\label{sol6}
\end{equation}
The solutions (\ref{sol1})-(\ref{sol6}) are characterized by $p$ and $s$ as free parameters. Also, as was defined in (\ref{variables1}), variable $y$ depends on arbitrary function $\omega(\phi)$. Bearing this in mind we can construct general solutions to the WDW equation by superposition of the given solutions. Later on we obtain the wormholes solutions for $k=1$.


\subsection{Case $\Lambda(y)=\Lambda_0$}

Now, we study the WDW equation with $\Lambda(y)=\Lambda_0$, where
$\Lambda_0$ is a constant, this case is equivalent to say that the 
effective cosmological term $\lambda(\phi)$ is proportional to the scalar field $\phi$, 
thus from Eq. (\ref{wdw1}) we have the following wave equation
\begin{equation}
\left[x^2\frac{\partial^2}{\partial x^2} + px\frac{\partial}{\partial x} - kx^4 + \Lambda_0x^6
- \frac{\partial^2}{\partial y^2}  \right]
\psi(x,y)=0.
\label{wdw3}
\end{equation}
Setting $\psi(x,y)=e^{isy}X(x)$, the equation for the variable $x$ is
\begin{equation}
x^2X^{\prime\prime}(x) + pxX^{\prime} +
(s^2 + \Lambda_0x^6 - kx^4)X = 0 ,
\label{eqX}
\end{equation}
once again we will consider different values of $k$ in this equation.
\begin{flushleft}
\noindent  {\em i) $k=\pm 1$} 
\end{flushleft}
For the closed  and hyperbolic model, the above equation is solvable by power series\cite{Wiltshire}, but the solution is a  more complicated, then in order to look for another expressed in terms of standard functions \cite{Abra}, we take $X(x)=z^Af(z) \quad (z=x^2)$, thus Eq. (\ref{eqX}) transform into
\begin{equation}
4z^2f^{\prime\prime}(z) + (2+p+8A)f^\prime +  
\left[4A^2 + (p-2)A + s^2 \mp z^2+\Lambda_0 z^3\right]f = 0,
\label{eqZ-1}
\end{equation}
in order to get an exact solution of this equation, we need to satisfy $2+p+8A=0$ and  $4A^2 + (p-2)A + s^2 = 0$ this imply that
$A=-\frac{2+p}{8}$ and  $p = \frac{-2\pm2\sqrt{2-s^2}}{3}$ then, Eq. (\ref{eqZ-1}) reduce to a more simple form
\begin{equation}
f^{\prime\prime}(z)+\frac{1}{4}\left(\Lambda_0 z \mp 1\right)f=0,
\label{eqZ-2}
\end{equation}
next, we use the substitution $f(z)=U(\varphi)$ with 
$\left(2\Lambda_0\varphi\right)^\frac{2}{3} = \Lambda_0 z \mp 1$, thus  
the universe wavefunction in terms of Airy functions is
\begin{eqnarray}
&&\psi(x,y) =  \left\{C_1Ai\left[-\left(2\Lambda_0\right)^
{-\frac{2}{3}}\left(\Lambda_0x^2 \mp 1\right)\right]
+ C_2Bi\left[-\left(2\Lambda_0\right)^{-\frac{2}{3}}
\left(\Lambda_0x^2 \mp 1\right)\right] 
 \right\}
{}\nonumber \\ 
&&x^{-2p - 1 + 2\sqrt{p^2-p - s^2 + 1/4}}
\left\{C_3e^{isy} + C_4e^{-isy}\right\}.
\label{sol8}
\end{eqnarray}
\begin{flushleft}
\noindent  {\em ii) $k=0$} 
\end{flushleft}
In a flat FRW universe, the wavefunction is
\begin{equation}
\psi_s(x,y) = x^{\frac{1-p}{2}}y^{\frac{1-q}{2}}
\left[C_1I_M\left(\frac{\sqrt{\Lambda_0}}{3}x^3\right) +
C_2I_{-M}\left(\frac{\sqrt{\Lambda_0}}{3}x^3\right)\right]
\left[C_3e^{isy} + C_4e^{-isy}\right],
\label{sol9}
\end{equation}
where $I_{\pm M}$ is a modified Bessel function and $M=\frac{1}{3}\sqrt{\left(\frac{1-p}{2}\right)^2-s^2}$, and by superposition we can obtain wavefunctions satisfying particular boundary conditions. On the other hand, for the case $k=\pm 1$ we do not have a complete set of solutions, but only two particular solutions for the specific choice of the factor ordering parameter $p$.


\section{ Gauge N=1/x }

In this section our aim is show another solvable quantum model with $N=1/x$, then the action (\ref{action3}) becomes
\begin{equation}
S=\int \frac{1}{2}\left[x^2x^{\prime2}-x^4y^{\prime2}+x^2\Lambda(y)-k\right]d\tau,
\label{action-5}
\end{equation}
and with the following change of independent variables
\begin{equation}
\alpha = x^2\cosh(2y), \quad \quad \beta = x^2\sinh(2y),
\label{variables2}
\end{equation}
the action (\ref{action-5}) takes the symmetric form
\begin{equation}
S=\frac{1}{2}\int\left[\frac{1}{4}\left(\alpha^{\prime 2} -
\beta^{\prime 2}\right) + \Lambda_1\alpha + \Lambda_2\beta - k
\right]\,d\tau,
\label{action6}
\end{equation}
where we have chosen the cosmological term as 
\begin{equation}
\Lambda(y) = \Lambda_1\cosh(2y)+\Lambda_2\sinh(2y),
\label{Lambda}
\end{equation}
where $\Lambda_1$ and $\Lambda_2$ are constants, and the canonical momenta are given by
\begin{equation}
\pi_\alpha=\frac{\alpha^\prime}{4},\quad\quad \pi_\beta=-\frac{\beta^\prime}{4}.
\label{momenta-2}
\end{equation}
The Hamiltonian is constructed as
\begin{equation}
{\cal H}=2\pi_\alpha^2-2\pi_\beta^2 + \frac{1}{2}(k-\Lambda_1\alpha-\Lambda_2\beta),
\label{Hamiltonian-2}
\end{equation}
thus the corresponding WDW equation is
\begin{equation}
\left[\frac{\partial^2}{\partial \alpha^2}  - 
\frac{\partial^2}{\partial \beta^2}  + \frac{1}{4}
\left(\Lambda_1\alpha + \Lambda_2\beta - k\right)\right]
\psi(\alpha,\beta) = 0.
\label{wdw4}
\end{equation} 
In particular, we can get exact solutions setting $\psi(\alpha, \beta) = A(\alpha)B(\beta)$ and introducing  the new variables 
\begin{equation}
\gamma=(2\Lambda_1)^{-\frac{2}{3}}(k-4s^2-\Lambda_1\alpha),\quad \quad
\eta=(2\Lambda_1)^{-\frac{2}{3}}(\Lambda_2\beta-4s^2),
\label{variables3}
\end{equation} 
where $s$ is the separation constant. Then we obtain Airy equations $G^{\prime\prime}(\gamma) 
- \gamma G=0$ and $H^{\prime\prime}(\eta) - \eta H=0$. Consequently, 
the universe wavefunction for this model is
\begin{eqnarray}
\psi(\alpha,\beta)=\biggl\{C_1 Ai\left[ (2\Lambda_1)^{-\frac{2}{3}}
(k-4s^2-\Lambda_1\alpha)\right] + 
C_2 Bi\left[ (2\Lambda_1)^
{-\frac{2}{3}}(k-4s^2-\Lambda_1\alpha)\right]\biggr\}
{}\nonumber \\ 
\biggl\{C_3 Ai\left[ (2\Lambda_2)^{-\frac{2}{3}}
(\Lambda_2\beta-4s^2)\right] + 
C_4 Bi\left[ (2\Lambda_2)^
{-\frac{2}{3}}(\Lambda_2\beta-4s^2)\right]\biggr\}.
\label{sol10}
\end{eqnarray}
To have a wavefunction for a particular theory we take the choice $\omega(\phi) =\omega_0=const.$, we obtain a cosmological term in the Brans-Dicke theory  that could lead to inflation in the classical theory\cite{Pimentel,Horiguchi}
\begin{equation}
\lambda(\phi)=\lambda_1\phi^{\rho+1} +  \lambda_2\phi^{-\rho+1},
\label{lambda1}
\end{equation}
where $\rho$ is given by $\rho=\sqrt{(2\omega_0+3)/3}$ then, variables $\alpha$ and $\beta$ becomes to
\begin{equation}
\alpha=\frac{a}{2}\left(\phi^{\rho+\frac{1}{2}}+\phi^{-\rho+\frac{1}{2}}
\right),\quad  \quad 
\beta=\frac{a}{2}\left(\phi^{\rho+\frac{1}{2}}-\phi^{-\rho+\frac{1}{2}}
\right),
\label{variables4}
\end{equation}
in this case of Brans-Dicke theory, the universe wavefunction is
\begin{eqnarray}
\psi(a,\phi)=\biggl\{
C_1Ai\left[(8\Lambda_1)^{-\frac{2}{3}}
\left(2k-8s^2-\Lambda_1 a\left(\phi^{\rho+\frac{1}{2}}+
\phi^{-\rho+\frac{1}{2}}\right)\right)\right]
\nonumber \\
+C_2Bi\left[(8\Lambda_1)^{-\frac{2}{3}}
\left(2k-8s^2-\Lambda_1 a\left(\phi^{\rho+\frac{1}{2}}+
\phi^{-\rho+\frac{1}{2}}\right)\right)\right]\biggr\}
\nonumber \\
\biggl\{C_3Ai\left[(8\Lambda_2)^{-\frac{2}{3}}
\left(\Lambda_2 a\left(\phi^{\rho+\frac{1}{2}}+
\phi^{-\rho+\frac{1}{2}}\right)-8s^2\right)\right]
{} \nonumber \\+
C_4Bi\left[(8\Lambda_2)^{-\frac{2}{3}}
\left(\Lambda_2 a\left(\phi^{\rho+\frac{1}{2}}+
\phi^{-\rho+\frac{1}{2}}\right)-8s^2\right)\right]
\biggr\}.
\label{sol11}
\end{eqnarray}
Depending on the sign of the argument the Airy functions are oscillatory or exponential, by superposition, different boundary conditions can be satisfied. In the next section we show some classical solutions for the general and some particular theories.


\section{Classical solutions}

The classical equations of motion for this model derived from the action (\ref{action6}) are
\begin{equation}
\alpha^{\prime\prime}(\tau)-2\Lambda_1=0,
\label{class1}
\end{equation}
\begin{equation}
\beta^{\prime\prime}(\tau)+2\Lambda_2=0,
\label{class2}
\end{equation}
with the Hamiltonian constraint equation given by
\begin{equation}
\alpha^{\prime2}-\beta^{\prime2}-4\Lambda_1\alpha-4\Lambda_2\beta+4k=0.
\label{class3}
\end{equation}
The solution of system (\ref{class1})-(\ref{class2}) is
\begin{equation}
\alpha(\tau)=\Lambda_1\tau^2+C_1\tau+C_2,
\label{solclass1}
\end{equation}
\begin{equation}
\beta(\tau)=-\Lambda_2\tau^2+C_3\tau+C_4,
\label{solclass2}
\end{equation}
where $C_1$ and $C_2$ are integration constants and satisfy $C_1^2-C_3^2-4\Lambda_1C_2-4\Lambda_2C_4+4k=0$. In the following we study simply choices of the $\omega(\phi)$ function, this lead us to Brans-Dicke and Barker theories, also we consider three general parameterized theories that have been studied by other authors\cite{Barrow1,Barrow2} in the Appendix. 


\subsection{Brans-Dicke theory}

Choosing $\omega(\phi)=\omega_0$, then $y=\frac{\rho}{2}\ln{\phi}$ and we obtain the following solution
\begin{eqnarray}
a(\tau)&=&\left[(\Lambda_1+\Lambda_2)\tau^2+(C_1-C_3)\tau+C_2-C_4\right]^m
\nonumber\\
&&\left[(\Lambda_1-\Lambda_2)\tau^2+(C_1+C_3)\tau+C_2+C_4\right]^{m-n},
\label{sol-gen-a}
\end{eqnarray}
\begin{equation}
\phi(\tau)=\left[\frac{(\Lambda_1-\Lambda_2)\tau^2+(C_1+C_3)\tau +C_2+C_4}
{(\Lambda_1+\Lambda_2)\tau^2+(C_1-C_3)\tau+C_2-C_4}\right]^{n},
\label{sol-gen-phi}
\end{equation}
where $m=(1+\rho)/4\rho$, $n=1/2\rho$ and $\rho=\sqrt{(2\omega_0+3)/3}$. Here we have a family of solutions, depending on several parameters, in the next subsection we take particular values of the parameter to have explicit solutions that are easier to analyze. We found singular and non singular solutions.


\subsubsection{ Non singular solution }

Setting $C_1=C_3=C_4=0, C_2=k/\Lambda_1$ and $\Lambda_1=\Lambda_2$ in the above equations, we have the solution
\begin{equation}
a(\tau)=a_0\left[\frac{\tau^2}{\tau_0^2}+1\right]^m,
\label{sol-ns-a}
\end{equation}
and
\begin{equation}
\phi(\tau)=\left[\frac{\tau^2}{\tau_0^2}+1\right]^{-n},
\label{sol-ns-phi}
\end{equation}
where $a_0=\left(k^2/\Lambda_1^2\right)^{\frac{1}{4}}$ and $\tau_0^2=k/2\Lambda_1^2$, this solution corresponds to a cosmological term of the form $\Lambda(\alpha,\beta)=\Lambda_1\sqrt{(\alpha+\beta)/(\alpha-\beta)}$ and is not singular for $k=1$, we can verify this, by direct substitution of the solution in the metric invariants
\begin{equation}
R=-6\left[\frac{k}{a^2}+2\phi^2 a^{\prime 2} + a\phi a^\prime\phi^\prime + a\phi^2 a^{\prime\prime}\right],
\label{Ricci-tau}
\end{equation}
\begin{equation}
R_1=\frac{3}{4a^4}\left[k-a^3\phi a^\prime\phi^\prime-a^3\phi^2 a^{\prime\prime}\right]^2,
\label{R1}
\end{equation}
\begin{equation}
R_2=-\frac{1}{\sqrt3}R_1^{3/2},
\label{R2}
\end{equation}
\begin{equation}
R_3=\frac{7}{\sqrt{12}}R_1^2.
\label{R3}
\end{equation}
Introducing the solutions (\ref{sol-ns-a}) and (\ref{sol-ns-phi}) in the Ricci scalar (\ref{Ricci-tau}) we obtain
\begin{eqnarray}
R&=&-\frac{6}{a_0^2\tau_0^4\left(\frac{\tau^2}{\tau_0^2}+1\right)^{2(m+1)}}
\Biggl[ k(\tau_0^2+\tau^4) + \nonumber \\
 &+&2\tau^2\Biggr\{k\tau^2_0+a_0^4m(6m-2n-1)\left(\frac{\tau^2}{\tau_0^2}+1\right)\Biggl\}+2a_0^4m\tau_0^2\left(\frac{\tau^2}{\tau_0^2}+1\right)\Bigg],
\label{Ricci-inv}
\end{eqnarray}
and the $R_1$ invariant takes the form
\begin{eqnarray}
R_1&=&-\frac{3}{4a_0^4\tau_0^8\left(\frac{\tau^2}{\tau_0^2}+1\right)^{4(m+1)}}
\Biggl[ k(\tau_0^4+\tau^4) + \nonumber \\
 &+&2\tau^2\left\{ k\tau_0^2 + 2a^4_0m(2n-2m+1)\left(\frac{\tau^2}{\tau_0^2}+1\right)\right\}-2a_0^4m\tau_0^2\left(\frac{\tau^2}{\tau_0^2}+1\right)\Biggr]^2.
\label{R1-inv}
\end{eqnarray}
It is easy to check that this two invariants are finite for all values of $\tau$.


\subsubsection{Singular solution}

Another classical cosmological solution can be obtained by choosing $C_1=C_3=0,\,\,C_2=C_4=k/2\Lambda_1\,\,$ and $\Lambda_1=\Lambda_2$, thus the solutions (\ref{sol-gen-a})-(\ref{sol-gen-phi}) are reduced to
\begin{equation}
a(\tau)=a_0^\prime\tau^{2m},
\end{equation}
\begin{equation}
\phi(\tau)=\phi_0^\prime\tau^{-2n},
\end{equation}
where $a_0^\prime=2^{\frac{1}{4}}\tau_0^{-\frac{n}{2}}$ and $\phi_0^\prime=\tau_0^{2n}$. These solutions in the cosmic time $t$ take the form
\begin{equation}
a(t)=a^\prime t^{\frac{2(1+\rho)}{2-3\rho}},
\end{equation}
\begin{equation}
\phi(t)=\phi^\prime t^{\frac{4}{3\rho-2}},
\end{equation} 
where $a^\prime=a_0^\prime\left\{ 2^{1/4}\tau^{3/4\rho}\left(\frac{1+\rho}{2\rho}\right)\right\}^{\frac{2(1+\rho)}{2-3\rho}}$ and $\phi^\prime=\phi_0^\prime\left\{ 2^{1/4}\tau^{3/4\rho}\left(\frac{1+\rho}{2\rho}\right)\right\}^{\frac{4}{3\rho-2}}$. For other theories see the Appendix.


\section{Quantum wormhole solutions}

In this part we show some special solutions of the WDW equation for $N=1$ and $N=1/x$, which are known as quantum wormholes.


\subsection{Case $\Lambda(y)=0$, $N=1$}

For the closed universe $(k=1)$, assuming $p=1$ the solution to WDW equation (\ref{wdw2}) is
\begin{equation}
\psi_s(x,y) = 
\left[C_1I_{is}\left(\frac{x^2}{2}\right) +
C_2K_{is}\left(\frac{x^2}{2}\right)\right]
\left[C_3e^{is y} + C_4e^{-is y}\right].
\label{sol2}
\end{equation}
If we choose the family of solutions
\begin{equation}
\psi_s(x,y) = Ce^{is y}K_{is}\left(\frac{x^2}{2}\right),
\label{sol3}
\end{equation}
then we can generates a quantum wormhole basis, namely, making wave superposition
\begin{equation}
\psi_{WH}(x,y)=\int_{-\infty}^{+\infty}C(s)e^{is y}
K_{is}\left(\frac{x^2}{2}\right)\,ds,
\label{sol4}
\end{equation}
if we take $C(s)=e^{i\mu s}$ $(\mu=const.)$, by means of a Kontorovich-Lebedev integral transform \cite{Bateman} we obtain
\begin{equation}
\psi_\mu(x,y) = e^{-\frac{x^2}{2}\cosh[2y + \mu]}.
\label{wormhole1}
\end{equation}
This solution is a quantum wormhole, because it satisfies the Hawking-Page regularity boundary conditions, i.e., the wavefunction is exponentially damped for large spatial geometry, and also, is regular when the spatial geometry degenerates \cite{Hawking1}. Wormholes may play an important role in solving problems associated with the complete evaporation of black holes, and it is believed that they produce effective interactions in the low energy physics \cite{Hawking2} that turn the coupling constants of nature into dynamical variables \cite{Coleman}. The set of wavefunctions $\psi_s$ and $\psi_\mu$ are spanning the same space of physical states, and are both bases of the Hilbert space of the model in the corresponding representation, and Eq. (\ref{sol4}) is the connection between these bases $\psi_s$ and $\psi_\mu$ \cite{Pimentel0,Garay}. For Euclidean wormholes in BW theory see \cite{Nandi}. 


\subsection{Case $\Lambda(y)=\Lambda_0$, $N=1$}

In a similar way, for a flat model $(k=0)$, we find for $\Lambda_0<0$  the following quantum wormhole basis
\begin{equation}
\psi_\mu(x,y)=e^{-\sqrt{\Lambda_0}x^3\cosh{[3y-\mu]}},
\label{wormhole2}
\end{equation}
again this solution satisfy the mentioned Hawking-Page wormhole  condition, and by direct substitution we can verify that basis (\ref{wormhole2}) is a particular solution of the Eq. (\ref{wdw3}).


\subsection{ Case $\Lambda(\eta/\xi)=0$, $N=1/x$}

Also quantum wormholes can be obtained from action (\ref{action3}) with $N=1/x$,  if we introduce the following independent variables
\begin{equation}
 \xi=x\cosh{y}, \qquad  \qquad \eta=x\sinh{y},
\label{variables5}
\end{equation}
then the new action is 
\begin{equation}
S=\frac{1}{2}\int \left[\xi^2\xi^{\prime2}-\xi^2\eta^{\prime2}-\eta^2\xi^{\prime2}+\eta^2\eta^{\prime2}-k+(\xi^2-\eta^2)^2\Lambda(\eta/\xi)\right]d\tau,
\label{action5}
\end{equation}
and the corresponding WDW equation is
\begin{equation}
\left[\frac{\partial^2}{\partial \xi^2}  - 
\frac{\partial^2}{\partial \eta^2} -  (\xi^2-\eta^2)k + (\xi^2-\eta^2)^2\Lambda(\eta/\xi )\right]
\psi(\xi,\eta) = 0.
\label{wdw6}
\end{equation}
Choosing $\Lambda(\eta/\xi)=0$ and considering a closed model $(k=1)$, Eq. (\ref{wdw6}) simplifies to
\begin{equation}
\left[\frac{\partial^2}{\partial \xi^2} - \xi^2 - 
\frac{\partial^2}{\partial \eta^2} + \eta^2  \right]
\psi(\xi,\eta) = 0,
\label{wdw7}
\end{equation}
we can see that this is, the equation for two harmonic oscillators with opposite signs of the energy with solution  
\begin{equation}
\psi(\xi,\eta)_{WH}=H_n(\xi)H_n(\eta)e^{-\frac{\xi^2+\eta^2}{2}}.
\label{wormhole3}
\end{equation}
where $H_n$ are Hermite polynomials of order $n$. These harmonic-oscillator solutions are a quantum wormhole basis, since satisfied the Hawking-Page regularity boundary condition, {\em (i)} the solutions are regular at the origin and {\em (ii)} damped at infinity.


\section{ Third quantization }

It is known that the WDW equation is a result of the quantization of a geometry and mater (second quantization of gravity). The procedure of quantizing the wavefunction $\psi$ is called third quantization\cite{Hosoya}, in this theory we consider $\psi$ as an operator acting on the state vectors of a system of universes and can be decomposed as
\begin{equation}
\psi(\alpha,\beta)=\int_{-\infty}^\infty ds\left[ C(s)\psi_s(\alpha,\beta) + C(s)^{\dag}\psi^*(\alpha,\beta)\right],
\label{int-0}
\end{equation}
where $\psi(\alpha,\beta)$ and $\psi^*(\alpha,\beta)$ form complete orthonormal sets of solutions to the WDW equation. This is in analogy with the quantum field theory, where $C(s)$ and $C^{\dag}(s)$ are creation and annihilation operators. Thus, we expect that the vacuum state in a third quantized theory is unstable and creation of universes from the
initial vacuum state takes place. In this view, the variable $\alpha$
plays the role of time, and variable $\beta$ the role of space. The wavefunction $\psi(\alpha,\beta)$ is interpreted as a quantum field in minisuperspace.

We assume that the creation and annihilation operators of universes obey the standard commutation relations
\begin{equation}
  [C(s), C^{\dag}(s^\prime)]= \delta(s-s^\prime),\quad \quad
  [C(s), C(s^\prime)]=[C^{\dag}(s), C^{\dag}(s^\prime)]=0,\quad
\label{commutation}
\end{equation}
and we expand the field $\psi$ in normal modes $\psi_s$ as shown in Eq. (\ref{int-0}). Here, the wave number $s$ is the momentum in Planck units and is very small. Let us consider now the last quantum model (\ref{wdw4}) for $\Lambda_2=0$ and closed universe $k=1$. Then, the WDW equation becomes
\begin{equation}
 \left[\frac{\partial^2}{\partial \alpha^2} - \frac{\partial^2}{\partial \beta^2} +
   \frac{1}{4}(\Lambda_1\alpha - 1)\right]
 \psi(\alpha,\beta) = 0.
\label{wdw5}
\end{equation}
A complete set of normalized positive frequency solutions to this equation \cite{Horiguchi} is
\begin{eqnarray}
\psi_s^+(\alpha,\beta)=\frac{e^{is\beta}}{(16\Lambda_1)^{\frac{1}{16}}}
\biggl\{Ai\left[ (2\Lambda_1)^{-\frac{2}{3}}
(1-4s^2-\Lambda_1\alpha)\right] + 
{}\nonumber \\
iBi\left[ (2\Lambda_1)^
{-\frac{2}{3}}(1-4s^2-\Lambda_1\alpha)\right]\biggr\},
\label{sol+}
\end{eqnarray}
and
\begin{eqnarray}
\psi_s^-(\alpha,\beta)=
\frac{\sqrt2\,e^{is\beta}}{(16\Lambda_1)^{\frac{1}{16}}}
\biggl\{e^{\frac{(1-4s^2)^{\frac{3}{2}}}{3\Lambda_1}}
Ai\left[ (2\Lambda_1)^{-\frac{2}{3}}(1-4s^2-\Lambda_1\alpha)\right]
{}\nonumber \\
+\frac{i}{2}e^{-\frac{(1-4s^2)^{\frac{3}{2}}}{3\Lambda_1}}
Bi\left[(2\Lambda_1)^{-\frac{2}{3}}(1-4s^2-\Lambda_1\alpha)\right]
\biggr\},
\label{sol-}
\end{eqnarray}
the wavefunctions $\psi_s^+$ and $\psi_s^-$ can be seen as a positive frequency out going and in modes, respectively, and these solutions are orthonormal with respect to the Klein-Gordon scalar product
\begin{equation}
\left<\psi_s,\psi_{s^\prime}\right>=
i\int \psi_s\stackrel{\leftrightarrow}{ \partial}_\beta
\psi_{s^\prime}^*d\beta = \delta(s-s^\prime).
\label{int-1}
\end{equation}
As both sets, (\ref{sol+}) and (\ref{sol-}) are complete, they are related to each other by the Bogoliubov transformation defined by
\begin{equation}
 \psi^+_s(\alpha,\beta)=
   \int dr\left[M^+(s,r)\psi_r^-(\alpha,\beta) +
   M^-(s,r)\psi_r^-(\alpha,\beta)\right].
 \label{int-2}
\end{equation}
The wavefunction $\psi(\alpha,\beta)$ can be expanded in terms of 
annihilation and creation operators for each set of modes, then
with wavefunctions (\ref{sol+}) and (\ref{sol-}) we obtain that the Bogoliubov coefficients $M^+(s,r)=\delta(s-r)v(s)$ and $M^-(s,r)=\delta(s+r)w(s)$ are
\begin{equation}
   M^{\pm}(s,r)=
      \delta(s\mp r)\frac{1}{\sqrt 2}\left[\frac{1}{2}
         e^{-\frac{(1-4s^2)^{\frac{3}{2}}}{3\Lambda_1}} +
      e^{\frac{(1-4s^2)^{\frac{3}{2}}}{3\Lambda_1}}\right].
\label{coeff}
\end{equation}         
The coefficients $M^\pm(s,r)$ are not equal to zero. Thus, two Fock spaces constructed with the help of the modes (\ref{sol+}) and (\ref{sol-}) are not equivalent and we have two different third quantized vacuum states (voids): The in-vacuum $\mid 0, in\rangle$ and out-vacuum $\mid 0, out\rangle$, (which are the states with no FRW-like universes) defined by
\begin{equation}
  C_{in}(s)\mid 0, in \rangle  = 0
    \quad \hbox{and} \quad
    C_{out}(s)\mid 0, out \rangle = 0,
  \label{states}
\end{equation}
where $s \in {\bf R}$. As was mentioned, since the vacuum states $\mid 0, in \rangle$ and $\mid 0, out \rangle$ are not equivalent, the birth of the universes from nothing may have place, where nothing is the vacuum state $\mid 0, in \rangle$. The average number of universes produced from nothing, in the s-th mode $N(s)$ is
\begin{equation}
  N(s)=\left<0,in\mid C_{out}^{\dag}(s)C_{out}(s)
  \mid 0, in \right>,
 \label{average-0}
\end{equation}
as follows from Eq. (\ref{coeff}) we get
\begin{equation}
N(s)=\frac{1}{2}\left[\frac{1}{2}
e^{-\frac{(1-4s^2)^{\frac{3}{2}}}{3\Lambda_1}} -
e^{\frac{(1-4s^2)^{\frac{3}{2}}}{3\Lambda_1}}\right]^2.
\label{average-2}
\end{equation}
In view of Coleman's wormhole mechanism \cite{Coleman} for the vanishing cosmological constant, we take the constraint $\Lambda_1 \le 1/8\pi \times 10^{-120}m_p^4$, also assuming $|s|<<1$, then notice that we can get the number of state $N(s)$  
\begin{equation}
N(s) \approx \frac{1}{2}e^{\frac{2}{3\Lambda_1}(1-4s^2)^{\frac{3}{2}}}.
\label{average-3}
\end{equation}
This result from third quantization shows that the number of the universes produced from nothing is exponentially large.


\section{Uncertainty relation}

In this section we will study third quantization for the gauge $N=1$, in order to get the uncertainty Heisenberg relation, we start from WDW equation (\ref{wdw1}), setting $\Lambda(y)=\Lambda_0$ we have
\begin{equation}
\left[x^2\frac{\partial^2}{\partial x^2} + px\frac{\partial}{\partial x} - \frac{\partial^2}{\partial y^2} 
- x^4(k - \Lambda_0x^2)\right]\psi(x,y)=0,
\label{wdw3q-1}
\end{equation}
if we choose $p=-1$ and introducing the new variable $z=x^2$, the above equation transform into
\begin{equation}
\left[\frac{\partial^2}{\partial z^2}-\frac{1}{4z^2}\frac{\partial^2}{\partial y^2}+
U(z)\right]\psi(z,y)=0,
\label{wdw3q-2}
\end{equation}
\begin{equation}
U(z)=-\frac{1}{4}(k-\Lambda_0z).
\label{pot-wdw-U}
\end{equation}
The third-quantized action to construct de WDW equation (\ref{wdw3q-2}) is
\begin{equation}
S=\frac{1}{2}\int\left[\left(\frac{\partial\psi}{\partial z}\right)^2-
\frac{1}{4z^2}\left(\frac{\partial\psi}{\partial y}\right)^2-
U(z)\psi^2\right]\,dz dy.
\label{action-3q-U}
\end{equation}
Now we  use a Fourier decomposition in order to obtain decoupled degrees of freedom, the universe field $\psi(z,y)$ will be expand in sine and cosine functions. We assume that our quantum system is confined to a one-dimensional box with periodic boundary conditions, with the coordinate length fixed at a arbitrary value $M$, then
\begin{equation}
\psi(z,y)=\frac{\sqrt{2\pi}}{M}\left\{\psi(z,0)+\sum_{q=2\pi n/M} \frac{1}{\sqrt{2}}
\left[\psi_+(z,q)\cos{qy}+\psi_-(z,q)\sin{qy}\right]\right\},
\label{Fourier}
\end{equation}
where $\psi_+(z,-q)=\psi_+(z,q)$ and $\psi_-(z,-q)=\psi_-(z,q)$. By substituting the expansion (\ref{Fourier}) into Eq. (\ref{action-3q-U}), we obtain that 
\begin{equation}
L=\frac{1}{2}\sum_\sigma\left[\left(\frac{\partial\psi_\sigma}{\partial z}\right)^2-\frac{q^2}{4z^2}\psi_\sigma^2-U(z)\psi_\sigma^2\right],
\label{Lagrangian3q-U}
\end{equation}
where we have denoted the mode variables $\psi(z,0)$ and $\psi_\pm(z,q)$ by $\psi_\sigma(z)$ and Eq. (\ref{Lagrangian3q-U}) is rescaled to $\psi_\sigma \to \sqrt{M/2\pi}\psi_\sigma$. The above sum includes zero mode $\psi(z,0)$ for each pair $(q,-q)$, in this way the mode variable $\psi_\sigma$ is completely decoupled from each other. Now the Hamiltonian is given by
\begin{equation}
H_\sigma=\sum_\sigma\frac{1}{2}\left[\pi_\sigma^2+\left(\frac{q^2}{4}+U(z)
\right)\psi_\sigma^2\right].
\label{Hamiltonian3q-U}
\end{equation}
To quantize this system (\ref{Hamiltonian3q-U}), we impose the equal time commutation relations
\begin{equation}
\left[\hat\psi_\sigma,\hat\pi_\sigma^\prime\right]=i\delta_{\sigma,
\sigma^\prime}.
\end{equation}
The wave functional of any energy eigenstate is factorized as
\begin{equation}
\Psi = \prod_\sigma\Psi_\sigma\left[z,\psi_\sigma\right].
\end{equation}
In order to get the Heisenberg uncertainty relation we use the Schr\"odinger picture, then we make the substitution $\hat\psi \to \psi_\sigma$ and $\hat\pi_\sigma \to -i\frac{d}{d\psi_\sigma}$. 
Then, the Schr\"odinger equation for each mode variable is
\begin{equation}
i\frac{\partial \Psi_\sigma}{\partial z} = \frac{1}{2}\left[-\frac{\partial^2}{\partial\psi_\sigma^2} + \left(\frac{q^2}{4z^2}+U(z)\right)\psi_\sigma^2\right]\Psi_\sigma.
\label{Sch}
\end{equation}
We will solve the above wave functional equation by using the Gaussian ansatz
\begin{equation}
\psi_\alpha[z,\psi_\sigma] = C\exp\{-\frac{1}{2}A_\alpha(z,q)[\psi_\sigma(z,q)-\theta_\sigma(z,q)]^2 + iP_\sigma(z,q)[\psi_\sigma(z,q)-\theta_\sigma(z,q)]\},
\label{ansatz}
\end{equation}
\begin{equation}
A_\alpha(z,q) = D_\sigma(z,q) + iF_\sigma(z,q),
\label{deff}
\end{equation}
where the real functions $D_\sigma(z,q)$, $F_\sigma(z,q)$ $P_\sigma(z,q)$, and $\theta_\sigma(z,q)$ have to be determined from Eq. (\ref{Sch}). $C$ is a normalization of the wave function. The inner product of two functionals $\psi_{1\sigma}(z,\psi_\sigma)$ and $\psi_{2\sigma}(z,\psi_\sigma)$ is defined by
\begin{equation}
<\psi_1\mid \psi_2>_z = \int d\psi_\sigma\psi_{1\sigma}[z,\psi_\sigma]
\psi^*_{2\sigma}[z,\psi_\sigma].
\label{inner2}
\end{equation}
We shall calculate Heisenberg's uncertainty relation, the dispersion of $\psi_\sigma$ is given by $\left(\Delta\psi_\sigma\right)^2 \equiv <\psi_\sigma^2>_z - <\psi_\sigma>^2_z,$ from Eqs. (\ref{ansatz}) and (\ref{inner2}) we have
\begin{equation}
<\psi_\sigma^2>_z = \frac{1}{2D_\sigma(z,q)} + \eta^2_\sigma(z,q), \quad
<\psi_\sigma>_z = \eta^2_\sigma(z,q),
\end{equation}
then 
\begin{equation}
\left(\Delta\psi_\sigma\right)^2 = \frac{1}{2D_\sigma(z,q)},
\end{equation}
and the dispersion of $\pi_\sigma$ is  $\left(\Delta\pi_\sigma\right)^2 \equiv  <\pi_\sigma^2>_z - <\pi_\sigma>^2_z$, where
\begin{equation}
<\pi_\sigma^2>_z = \frac{D_\sigma(z,q)}{2} + 
\frac{F^2_\sigma(z,q)}{2D_\sigma(z,q)} + P^2_\sigma(z,q), \quad
<\pi_\sigma>_z = P_\sigma(z,q),
\end{equation}
then simplifying
\begin{equation}
\left(\Delta\pi_\sigma\right)^2 = \frac{D_\sigma(z,q)}{2} + 
\frac{F^2_\sigma(z,q)}{2D_\sigma(z,q)},
\end{equation}
thus we obtain the uncertainty relation
\begin{equation}
\left(\Delta\psi_\sigma\right)^2\left(\Delta\pi_\sigma\right)^2 = \frac{1}{4}
\left\{1+\left[\frac{F_\sigma(z,q)}{D_\sigma(z,q)}\right]^2\right\}.
\label{rel-uncertainty}
\end{equation}
In order to obtain $ F_\sigma(z,q)$ and $D_\sigma(z,q)$ we substitute the ansatz (\ref{ansatz}) into Eq. (\ref{Sch}), thus we obtain the equation of motion for $A_\sigma(z,q)$:
\begin{equation}
-i\frac{d}{dz}A_\sigma(z,q) = -A^2_\sigma(z,q) + \frac{q^2}{4z^2} + U(z).
\end{equation}
Now we substitute
\begin{equation}
A_\sigma(z,q) = -i\frac{d}{dz}\ln{u_\sigma(z,q)} ,
\label{mot}
\end{equation}
where $u_\sigma(z,q)$ is the solution of the WDW equation
\begin{equation}
\left[\frac{\partial^2}{\partial z^2} + \frac{q^2}{4z^2} + U(z)\right]u_\sigma(z,q) = 0.
\label{wdw3q-3}
\end{equation}
Now  first, we analyze the behavior of $u_\sigma$ for large scales $(a\to\infty)$, then the second term  $q^2/4z^2$ in the wave equation (\ref{wdw3q-3}) can be neglected, and its solution in terms of Airy functions is
\begin{equation}
u_\sigma(z,q)=\hbox{Ai}\left[\left(4\Lambda_0^2\right)^{-\frac{1}{3}}(\Lambda_0 z-k)\right] + \varrho\hbox{Bi}\left[\left(4\Lambda_0^2\right)^{-\frac{1}{3}}(\Lambda_0 z-k)\right] ,
\label{wdw-2-solut}
\end{equation}
where $\varrho$ is a complex constant. Substituting (\ref{wdw-2-solut}) into  (\ref{mot}), we obtain that
\begin{equation}
A_\sigma(z,q)=i\Lambda_0(4\Lambda_0^2)^{\frac{1}{3}} 
\frac{\hbox{Ai}\left[\left(4\Lambda_0^2\right)^{-\frac{1}{3}}(\Lambda_0 z-k)\right] + \varrho\hbox{Bi}^\prime\left[\left(4\Lambda_0^2\right)^{-\frac{1}{3}}(\Lambda_0 z-k)\right]}
{\hbox{Ai}\left[\left(4\Lambda_0^2\right)^{-\frac{1}{3}}(\Lambda_0 z-k)\right] + \varrho\hbox{Bi}\left[\left(4\Lambda_0^2\right)^{-\frac{1}{3}}(\Lambda_0 z-k)\right]},
\label{eq-A}
\end{equation}
where the prime denotes differentiation with respect to $z$, also we have
\begin{equation}
D_\sigma(z,q)=\hbox{Re}A_\sigma(z,q)= 
\Lambda_0(4\Lambda_0^2)^{-\frac{1}{3}} 
\frac{(\hbox{Im} \varrho)(\hbox{Ai}^\prime + \varrho\hbox{Bi}^\prime)}
{\mid \hbox{Ai} + \varrho\hbox{Bi} \mid^2},
\end{equation}
and
\begin{equation}
F_\sigma(z,q)=-\hbox{Im}A_\sigma(z,q)=
\Lambda_0(4\Lambda_0^2)^{-\frac{1}{3}} 
\frac{\hbox{Ai}^\prime\hbox{Ai} + \mid\varrho\mid^2\hbox{Bi}^\prime\hbox{Bi} + (\hbox{Re}\varrho)(\hbox{Bi}^\prime\hbox{Ai}+\hbox{Ai}^\prime\hbox{Bi})}
{\mid \hbox{Ai} + \varrho\hbox{Bi} \mid^2},
\end{equation}
therefore the uncertainty relation becomes
\begin{equation}
\left(\Delta\psi_\sigma\right)^2\left(\Delta\pi_\sigma\right)^2 = \frac{1}{4}
\left\{1+\frac{\pi^2}{(\hbox{Im}\varrho)^2}\left[\hbox{Ai}^\prime\hbox{Ai} + \mid\varrho\mid^2\hbox{Bi}^\prime\hbox{Bi} + (\hbox{Re}\varrho)(\hbox{Bi}^\prime\hbox{Ai}+\hbox{Ai}^\prime\hbox{Bi})\right]^2\right\}.
\label{end-uncertainty-1}
\end{equation}
For large scales $(a\to\infty)$, the above uncertainty relation is given by
\begin{eqnarray}
\left(\Delta\psi_\sigma\right)^2\left(\Delta\pi_\sigma\right)^2 &\simeq& 
\frac{1}{4}
\Biggl\{1+\frac{1}{(\hbox{Im}\varrho)^2}\Biggl[(\mid\varrho\mid^2-1)\sin{\biggl[\frac{4}{3}(4\Lambda_0^2)^{-\frac{3}{4}}(k-\Lambda_0z)^{\frac{3}{2}}+\frac{\pi}{2}\biggr]} \nonumber \\
&-&2(\hbox{Re}\varrho)\cos{\biggl[\frac{2}{3}(4\Lambda_0^2)^{-\frac{3}{4}}(k-\Lambda_0z)^{\frac{3}{2}}+\frac{\pi}{2}\biggr]}\Biggr]^2\Bigg\},
\label{uncertainty-1}
\end{eqnarray}
for $\varrho=-i$, we get that 
\begin{equation}
\left(\Delta\psi_\sigma\right)^2\left(\Delta\pi_\sigma\right)^2 \simeq \frac{1}{4},
\label{uncertainty-2a}
\end{equation}
and when $\varrho\neq-i$ and $\mid \beta \mid >> 1$
\begin{equation}
\left(\Delta\psi_\sigma\right)^2\left(\Delta\pi_\sigma\right)^2 \simeq \frac{1}{4}\left[1+O(\mid\beta\mid^2)\right].
\label{uncertainty-2b}
\end{equation}
This means that the quantum fluctuations of the third-quantized closed universe field are bounded at a finite value according Eq. (\ref{end-uncertainty-1}), during the universe expansion $(a\to \infty)$. Now we study the behavior of $u_\sigma$ for small scales $(a\to 0)$. Assuming that $0<\Lambda_0\leq 1$ in Planck units and $k=1$, the solution to Eq. (\ref{wdw3q-3}) is
\begin{equation}
u_\sigma(z,q)=\sqrt{z}\left[I_\nu\left(\frac{z}{2}\right)+\gamma K_\nu\left(\frac{z}{2}\right)\right],
\label{wdw-2-sol}
\end{equation}
where $I_\nu$ and $K_\nu$ are modified Bessel functions, $\gamma$ is a complex constant, and $\nu=\frac{1}{4}\sqrt{1-q^2}$. Substituting the general solution (\ref{wdw-2-sol}) into Eq. (\ref{mot}), we obtain
\begin{equation}
D_\sigma=\hbox{Re}A_\sigma=\frac{\hbox{Im}\gamma(K_\nu^\prime I_\nu-I_\nu^\prime K_\nu)}{2\mid I_\nu+\gamma K_\nu\mid^2},
\end{equation}
\begin{equation}
F_\sigma=\hbox{Im}A_\sigma 
= -\frac{z^{-1}\mid I_\nu+\gamma K_\nu\mid^2 + I^\prime_\nu I_\nu+
\mid\gamma\mid^2K^\prime_\nu K_\nu + \hbox{Re}\gamma(K^\prime_\nu I_\nu
+I^\prime_\nu K_\nu)}{2\mid I_\nu+\gamma K_\nu\mid^2},
\end{equation}
the prime denotes differentiation with respect to $z$. Then, the uncertainty relation is
\begin{equation}
(\Delta\psi_\sigma)^2(\Delta\pi_\sigma)^2=
\frac{1}{4}\left\{1+G_\sigma^2\right\},
\label{uncert-2f}
\end{equation}
where
\begin{equation}
G_\sigma=(\hbox{Im}\gamma)^{-1}z^2\left[z^{-1}(I_\nu^2+\mid\gamma\mid^2 K_\nu^2) + I^\prime_\nu I_\nu+
\mid\gamma\mid^2K^\prime_\nu K_\nu + \hbox{Re}\gamma(K^\prime_\nu I_\nu
+I^\prime_\nu K_\nu) \right],
\end{equation}
since we have assumed that our system is confined to a one-dimensional box with periodic boundary conditions, with the coordinate length fixed at an arbitrary value $M$, if we take the limit $M \to \infty$, then we can choose  $\nu^2>0$, and for small scales $(a\to 0)$ the asymptotic behavior\cite{Abra} of Eq.  (\ref{uncert-2f}) is
\begin{equation}
(\Delta\psi_\sigma)^2(\Delta\pi_\sigma)^2=
\frac{1}{4}\left\{1+Qz^{-4\mid\nu\mid}\right\},
\end{equation}
where $Q$ is some positive constant. This means that the fluctuation of the third-quantized universe field becomes large for small scales $(a\to 0)$. For the case in which $\nu^2<0$, so that $\nu$ is pure imaginary, we obtained an oscillatory behavior of $G_\sigma$ and does not have a definite magnitude\cite{Horiguchi-2} when $a \to 0$. 


\section{ Conclusions and discussion }

In this paper we have investigated some issues of quantum cosmology of simple minisuperspace models in the Bergmann-Wagoner theory, after quantizing the models we have obtained a class of separable Wheeler-DeWitt equations. We construct universe wavefunctions for flat, closed and open cosmologies.  

The first physical consideration of wavefunctions here obtained, is related with the fact of that in quantum gravity the topology of spacetime is expected to fluctuate on Planck scales\cite{Wheeler}. The spacetime might be multiply connected, these connections are microscopic wormholes. Hawking and Page\cite{Hawking1} have argued that wormholes are  solutions of the WDW equation, which are exponentially damped for large three-geometries, and regular in some suitable way when the three-geometry collapses to zero. We have shown two discrete basis of wormholes for $\lambda(\phi)=0$ and $\lambda(\phi)=3\Lambda_0\phi$ cosmological terms. Also, we have presented  singular and non singular general solutions in the closed Brans-Dicke cosmology.

The second physical consideration is related with the third quantization of the universe. The solutions of WDW equation have problems with the probabilistic interpretation. In the usual formulation of quantum mechanics a conserved positive-definite probability density is required for a consistent interpretation of the physical properties of a given system, and the universe in the quantum cosmology perspective, does not satisfy this requirement, because the WDW equation is a hyperbolic second order differential equation, there is no conserved positive-definite probability density as in the case of the Klein-Gordon equation, an alternative to this, is to regard the wavefunction as a quantum field in minisuperspace rather than a state amplitude, and the strategy is to perform a third quantization in analogy with the quantum field theory. Thus, the aim of this approach is to construct a consistent probabilistic measure in quantum gravity by promoting the wavefunction of the universe to a quantum field operator that acts on a Hilbert space of states. The vacuum state in this space is identified as the state where the universe does not exist. Topology changing processes can then be described by including self-interaction of the universe field. In quantum field theory, the particles are created from the vacuum by a time-varying external potential and this suggests that universes could be created via a similar process. In the third quantization the universe field is expanded into positive frequency in- and out-mode functions and their hermitian conjugates. The in- and out-modes are related to one another by the Bogoliubov coefficients and these determine the number of universes in a given mode. We have found for a simple minisuperspace model with a potential term $\lambda(y)=\Lambda_1\cosh{(2y)}$, that the number of the universes produced from nothing is exponentially large.

On the other hand, quantum fluctuations of flat and open minisuperspace models have been analyzed by the method of a time dependent Hermitian invariant\cite{Abe}, and was obtained that the fluctuation around each WDW trajectory converges to its minimum rapidly in the course of the cosmic expansion. Nevertheless, the closed models have not been analyzed by this method. We studied the third-quantization of a closed minisuperspace model by using the functional Schr\"odinger equation in order to investigate the Heisenberg uncertainty relation, and we found that quantum fluctuation of the third-quantized universe field becomes small for large scales $(a\to\infty)$ in the course of cosmic expansion. Also we found an exponentially large dominance of quantum effects for small scales $(a\to 0)$. The opposite  behavior for the uncertainties was obtained by Phole\cite{Pohle}, the difference is do to the fact that he was considering the classically forbidden region as was pointed out by Horiguchi\cite{Horiguchi-2}.
 

\section{Acknowledgments}


C Mora was supported by a  COFAA-EDD-Instituto Polit\'ecnico Nacional Grant.


\appendix

\section{Classical solutions for some Scalar-Tensor theories}
\label{app:cs}

Here we consider some particular choices for the coupling function $\omega (\phi)$, that have been used previously by some other authors. A common property of these coupling functions is that they give an analytical result for the integral in Eq.(\ref{variables1}).


\subsection{ Barker's theory }

The choice $\omega({\phi})=\frac{4-3\phi}{2\phi-2}$ lead us to the Barker's theory. This coupling function has the characteristic of making $\dot G= 0$ to first order in the weak-field limit \cite{Barker}. Using this coupling function we obtain the solution
\begin{eqnarray}
a(\tau)&=&\left\{\tan{\ln\left[\frac{(\Lambda_1+\Lambda_2)\tau^2+(C_1-C_3)\tau+C_2-C_4}
{(\Lambda_1-\Lambda_2)\tau^2+(C_1+C_3)\tau+C_2+C_4}\right]^{\sqrt{3}} }
+ 1\right\}^{-\frac{1}{2}} \nonumber \\
&& 
\lbrack\lbrack (\Lambda_1+\Lambda_2)\tau^2+(C_1-C_3)\tau+C_2-C_4
\rbrack \nonumber \\
&&\lbrack (\Lambda_1-\Lambda_2)\tau^2+(C_1+C_3)\tau+C_2+C_4 
\rbrack\rbrack^\frac{1}{4},
\end{eqnarray}
and
\begin{equation}
\phi(\tau) = \tan\ln{\left[\frac{(\Lambda_1+\Lambda_2)\tau^2+(C_1-C_3)\tau+C_2-C_4}
{(\Lambda_1-\Lambda_2)\tau^2+(C_1+C_3)\tau+C_2+C_4}\right]^{\sqrt3}}
+ 1.
\end{equation}


\subsection{ Theory 1 }

For the coupling function $2\omega(\phi)+3=B_1^2\phi^{2(l+1)}$, $B_1>0$ \cite{Barrow1}, 
we obtain the following solution
\begin{eqnarray}
a(\tau)&=&\left\{\ln\left[\frac{(\Lambda_1+\Lambda_2)\tau^2+(C_1-C_3)\tau+C_2-C_4}
{(\Lambda_1-\Lambda_2)\tau^2+(C_1+C_3)\tau+C_2+C_4}\right]^{\frac{\sqrt{3}(l+1)}{2B_1}}
\right\}^{-\frac{1}{2(l+1)}}\nonumber \\
&& \lbrack\lbrack (\Lambda_1+\Lambda_2)\tau^2+(C_1-C_3)\tau+C_2-C_4
\rbrack \nonumber \\
&&\lbrack (\Lambda_1-\Lambda_2)\tau^2+(C_1+C_3)\tau+C_2+C_4 
\rbrack\rbrack^\frac{1}{4},
\end{eqnarray}
\begin{equation}
\phi(\tau)=\left\{\ln\left[\frac{(\Lambda_1+\Lambda_2)\tau^2+(C_1-C_3)\tau+C_2-C_4}
{(\Lambda_1-\Lambda_2)\tau^2+(C_1+C_3)\tau+C_2+C_4}\right]^{\frac{\sqrt{3}(l+1)}{2B_1}}
\right\}^{\frac{1}{l+1}}.
\end{equation}


\subsection{ Theory 2 }

If we take $2\omega(\phi)+3=B_2\mid \ln\left(\frac{\phi}{\phi_0}\right)\mid^{-2\delta}$ \cite{Barrow2}, the resulting solution to the cosmological equations is given by


\subsubsection{$\delta \neq 1$}

\begin{eqnarray}
&a&(\tau)=\phi_0^{-\frac{1}{2}}\exp\left\{-\frac{1}{2}
\left\{
\sqrt{\frac{3}{B_2}}
(1-\delta)
\,\ln
\left[
\frac{(\Lambda_1+\Lambda_2)\tau^2+(C_1-C_3)\tau+C_2-C_4}{(\Lambda_1-\Lambda_2)\tau^2+(C_1+C_3)\tau+C_2+C_4}
\right]
  \right\}^{\frac{1}{1-\delta}}\right\}
\nonumber \\
&& \left\{\lbrack (\Lambda_1+\Lambda_2)\tau^2+(C_1-C_3)\tau+C_2-C_4
\rbrack
\lbrack (\Lambda_1-\Lambda_2)\tau^2+(C_1+C_3)\tau+C_2+C_4 
\rbrack\right\}^\frac{1}{4},
\end{eqnarray}
and
\begin{equation}
\phi(\tau)=\phi_0\exp{\left\{
\sqrt{\frac{3}{B_2}}(1-\delta)\,\ln\left[\frac{(\Lambda_1+\Lambda_2)\tau^2+(C_1-C_3)\tau+C_2-C_4}{(\Lambda_1-\Lambda_2)\tau^2+(C_1+C_3)\tau+C_2+C_4} \right]
  \right\}}^{\frac{1}{1-\delta}}.
\end{equation}


\subsubsection{ $\delta = 1$ }

\begin{eqnarray}
&a&(\tau)=\phi_0^{-\frac{1}{2}}\exp\left\{-\frac{1}{2} 
\left[
 \frac{(\Lambda_1+\Lambda_2)\tau^2+(C_1-C_3)\tau+C_2-C_4}{(\Lambda_1-\Lambda_2)\tau^2+(C_1+C_3)\tau+C_2+C_4}
\right]
^{\frac{1}{4\sqrt{3B_2}}}\right\}\nonumber \\
&& \left\{\lbrack (\Lambda_1+\Lambda_2)\tau^2+(C_1-C_3)\tau+C_2-C_4
\rbrack
\lbrack (\Lambda_1-\Lambda_2)\tau^2+(C_1+C_3)\tau+C_2+C_4 
\rbrack\right\}^\frac{1}{4} ,
\end{eqnarray}
and
\begin{equation}
\phi(\tau)=\phi_0\exp\left[
\frac{(\Lambda_1+\Lambda_2)\tau^2+(C_1-C_3)\tau+C_2-C_4}{(\Lambda_1-\Lambda_2)\tau^2+(C_1+C_3)\tau+C_2+C_4}  \right]^{\frac{1}{4\sqrt{3B_2}}}.
\end{equation}


\subsection{ Theory 3 }

For the last function considered here we choose  $2\omega(\phi)+3=\frac{B_3}{1-\left(\frac{\phi}{\phi_0}\right)^A} $ with $A>0$, $B_3>0$ \cite{Barrow2}, and  we obtain
\begin{eqnarray}
&a&(\tau)=\phi_0^{-\frac{1}{2}}
\left\{1-\hbox{tanh}\ln\left[\frac{(\Lambda_1+\Lambda_2)\tau^2+(C_1-C_3)\tau+C_2-C_4}
{(\Lambda_1-\Lambda_2)\tau^2+(C_1+C_3)\tau+C_2+C_4}\right]^{\frac{A}{2}\sqrt{\frac{3}{B_3}}}
  \right\}^{-\frac{1}{2A}} 
\nonumber \\
&& \left\{\lbrack (\Lambda_1+\Lambda_2)\tau^2+(C_1-C_3)\tau+C_2-C_4
\rbrack
\lbrack (\Lambda_1-\Lambda_2)\tau^2+(C_1+C_3)\tau+C_2+C_4 
\rbrack\right\}^\frac{1}{4},
\end{eqnarray}
and
\begin{equation}
\phi(\tau)=\phi_0
\left\{1-\hbox{tanh}\ln\left[\frac{(\Lambda_1+\Lambda_2)\tau^2+(C_1-C_3)\tau+C_2-C_4}
{(\Lambda_1-\Lambda_2)\tau^2+(C_1+C_3)\tau+C_2+C_4}\right]^{\frac{A}{2}\sqrt{\frac{3}{B_3}}}
  \right\}^{\frac{1}{A}} .
\end{equation}



\end{document}